\RequirePackage{fix-cm}
\documentclass[twocolumn]{svjour3}          % twocolumn
\smartqed  % flush right qed marks, e.g. at end of proof
\usepackage{graphicx}
\usepackage{soul}
\usepackage{color}
\begin{document}

\title{Effect of Boundary Constraints on the Nonlinear Flapping of Filaments Animated by Follower Forces
%Effect of Drag and Pre-Stress on Nonlinear Flapping Dynamics of Constrained Filaments Caused by Follower Forces%\thanks{Grants or other notes
}


\author{Soheil Fatehiboroujeni$^*$        \and
	Arvind Gopinath		\and
        Sachin Goyal %etc.
}


\institute{Soheil Fatehiboroujeni (Corresponding Author) \at
              Purdue University\\
              School of Engineering\\
              West Lafayette, IN, 47907\\
              Tel.: +1 209-291-9749\\
              \email{sfatehib@purdue.edu}           %  \\
           \and
           Arvind Gopinath \at
              University of California, Merced\\
              Department of Bioengineering\\
              Merced, CA, 95343\\
              \email{agopinath@ucmerced.edu}                          
            \and
            Sachin Goyal \at
              University of California, Merced\\
              Department of Mechanical Engineering\\
              Health Science Research Institute\\
              Merced, CA, 95343\\
              \email{sgoyal2@ucmerced.edu}
}

\date{Received: date / Accepted: date}

\maketitle

\begin{abstract}
Elastically driven filaments subjected to animating compressive follower forces provide a synthetic way to mimic oscillatory beating of active biological filaments such as eukaryotic cilia. The dynamics of such active filaments can, under favorable conditions, exhibit stable time-periodic responses that result due to the interplay of elastic buckling instabilities, geometric constraints, boundary conditions, and dissipation due to fluid drag. In this paper, we use a continuum elastic rod model to estimate the critical follower force required for onset of the stable time-periodic flapping oscillations in pre-stressed rods subjected to fluid drag. The pre-stress is generated by imposing either clamped-clamped or clamped-pinned boundary constraints and the results are compared with those of clamped-free case, which is without pre-stress . We find that the critical value increases with the initial slack--that quantifies the pre-stress, and strongly depends on the type of the constraints at the boundaries. The frequency of oscillations far from onset, however depends primarily on the magnitude of the follower force, not on the boundary constraints. Interestingly, oscillations for the clamped-pinned case are observed only when the follower forces are directed towards the clamped end. This finding can be exploited to design a mechanical switch to initiate or quench the oscillations by reversing the direction of the follower force or altering the boundary conditions.

%Active Filaments, Follower Load, Instability, Buckling
\end{abstract}

\section{Introduction}
\label{sec:1}

Cilia and flagella are micron-sized filamentous organelles found in eukaryotic cells that play a crucial role in biologically important processes such as locomotion, mucus clearance, embryogenesis  and cell motility \cite{Ainsworth,sleigh}. While the biophysical and biochemical mechanisms governing and regulating the activity of these oscillations are still not well understood, there is a growing interest in biomimetic applications of these structures in the field of microfluidics and soft robotics. For example, elastically connected beads actuated by time-periodic magnetic fields \cite{Drey,baba} or external chemical gradient \cite{sasaki,patteson,chelakkot} enable directed transport of cargo.

An alternate mechanism motivated by biological motor filament assays that can also yield controllable oscillations in this case driven by mechanical instabilities involves slender filaments subjected to follower forces. In passive contexts, nonconservative follower loads acting as either a point force or a distributed load play a crucial role in several contexts such as pipes conveying fluid \cite{pai1,pai2}, self-propelled structures \cite{wood} and flutter in rockets \cite{pai3}. Practical engineering applications of systems subjected to follower forces, as well as the theoretical developments on stability analysis of such systems are presented in a number of review studies \cite{Elish,Inger,Bolo,Langthjem}. In biological contexts, follower forces are realized in motor-filament aggregates inside cells and in-vitro assays wherein polar molecular motors attach to filaments and exert forces along their backbone \cite{MBC}.  

Mechanical responses such as buckling are a common motif in biology 
\cite{Ashkan,Robison,Gopinath11} where filaments are typically constrained in some way by their surroundings. Similar constraints to unfettered motion is seen in animated filaments. 
Building on these ideas, several researchers have effectively used continuum models for analyzing the post-buckling behavior of slender inextensible active or inactive filaments subjected to follower forces \cite{qin,de,kevin}. These studies have focused on the dynamics of free-free, fixed-free, and pinned-free filaments with the base state being a straight non-stressed filament. However, the role of pre-stress in emergent oscillations driven by distributed follower forces has received attention only recently despite the richness of its potential for bio-inspired applications \cite{sfb2,Bayly}.

In this paper, we study the effect of boundary constraints on the flapping oscillations of pre-stressed rods animated by follower forces. In particular, we focus on two statically indeterminate scenarios of boundary constraints: fixed-fixed (FF) and pinned-fixed (PF). By 'fixed-fixed', we refer to a rod clamped at both ends, and by 'pinned-fixed' we refer to a rod clamped at one end and attached to a pin joint at the other end allowing free rotation. In both scenarios, the rod is pre-stressed by decreasing the end-to-end distance, thereby generating a buckled shape, and then it is subjected to a uniformly distributed follower force along the centerline tangent. The lack of constraint at the free-end of a cantilever (fixed-free scenario) allows for either lateral oscillations or steady rotations to develop \cite{chelakkot}. In contrast, in statically indeterminate fixed-fixed (FF) and pinned-fixed (PF) scenarios, the slack generated by initial compression offers the necessary degree of freedom to allow for in-plane oscillations (flapping). The rod model presented here is three-dimensional, but due to planar perturbations and loads, the resulting oscillations remain planar. That said, we have observed that out-of-plane oscillations can emerge from two-dimensional base states. In this paper, however, we focus our attention exclusively on planar dynamics and will report our findings on three-dimensional oscillations in a follow-up publication. A broader impact of the results presented in this paper lies in recognizing how the interplay of geometry, elasticity, dissipation and activity unique to the pre-stressed scenarios can be harvested to move or manipulate fluid at various length scales.

\section{Model}
\label{sec:2}
%We consider a rod that is straight in the stress-free state. By moving one end of the rod towards the other and forcing the rod to bend due to bucking as shown in Figure 1, we generate pre-stress in the rod. Thus, pre-stress is controlled by the end-to-end length of the rod, $L_{ee} < L$. Here the slack $1-L_{ee} / L$ serves as a parameter characterizing the deviation form the stress free state.

\begin{figure*}
\begin{center}
  \includegraphics[width=1.72\columnwidth]{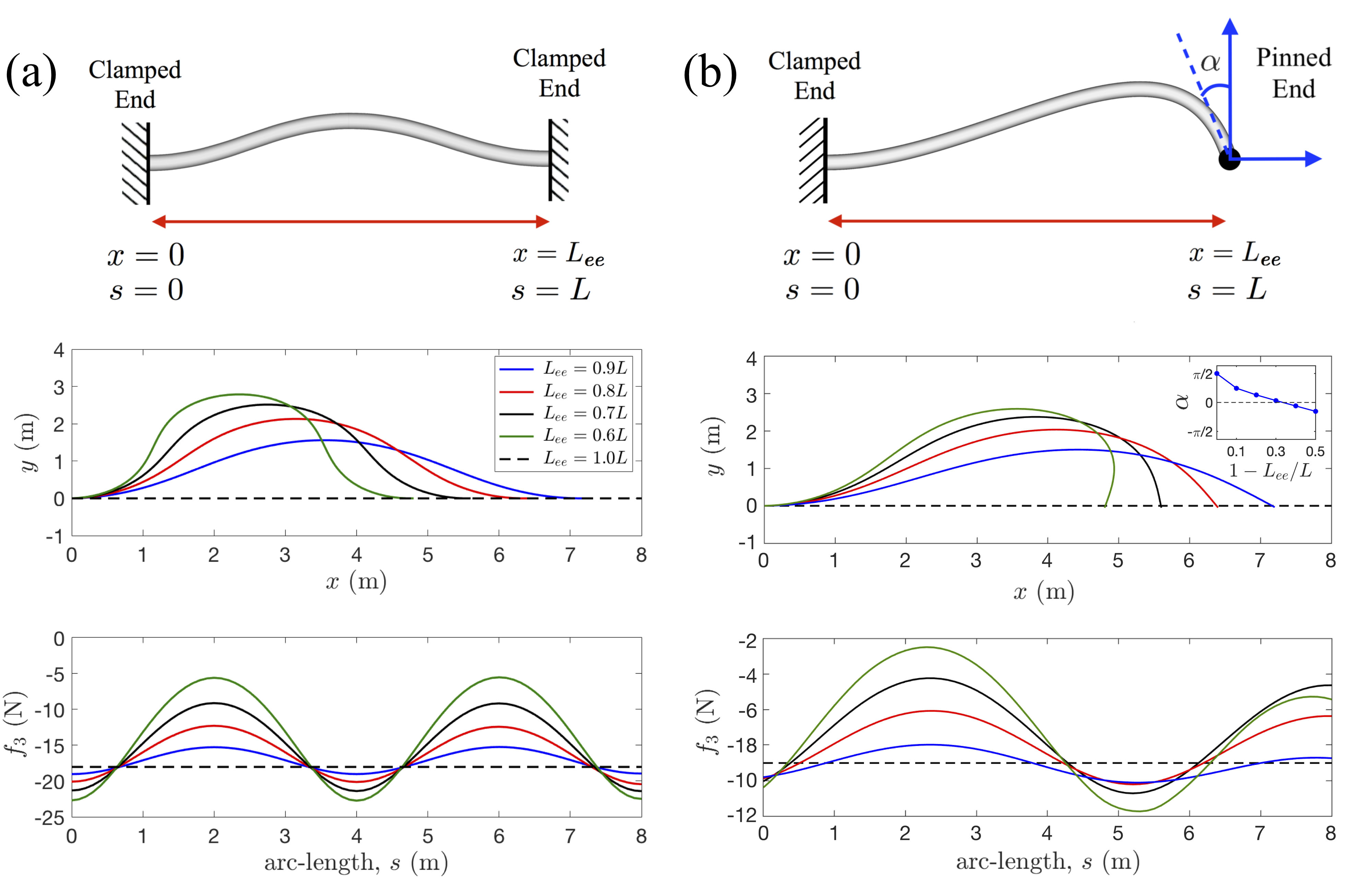}
\caption{The graphics on the top show the schematic representations of a rod of unstressed length $L$ in fixed-fixed (FF) scenario (a) and in pinned-fixed (PF) scenario (b). The end-to-end distance in the buckled state is $L_{\mathrm{ee}} < L$. The corresponding shapes of the rod centerline in the buckled states are shown in the middle row for different values of $L_{\mathrm{ee}}/L$. The dashed line corresponds to the {unbuckled} case $L_{\mathrm{ee}}/L=1.0$. Pre-stress contributes to the tension along the filament, $f_3$ (bottom row). For the shapes we study in this paper, there is a one-to-one correspondence between the tension, $f_3$, and pre-stress.}
\label{fig:1}   
\end{center}    
\end{figure*}

The continuum rod model that we use follows the classical approach of Kirchhoff \cite{kirk}, which assumes each cross-section of the rod to be rigid. The model is described in detail in \cite{sfb2}. To briefly summarize, equilibrium equations (\ref{linear_momentum}) and (\ref{angular_momentum}), and the compatibility conditions (\ref{position_continuity}) and (\ref{orient_continuity}) are given below:
\begin{equation}
m(\frac{\partial \vec{v}}{\partial t} + \vec{\omega} \times \vec{v}) - (\frac{\partial \vec{f}}{\partial s} + \vec{\kappa} \times \vec{f}) - \vec{F} = \vec{0} \label{linear_momentum}
\end{equation}
\begin{equation}
\underline{\mathbf{I_m}}\frac{\partial \vec{\omega}}{\partial t} + \vec{\omega} \times {{ \underline{\mathbf{I_m}}}}\vec{\omega} -(\frac{\partial \vec{q}}{\partial s} + \vec{\kappa} \times \vec{q}) + \vec{f} \times \vec{r} - \vec{Q}= \vec{0} \label{angular_momentum}
\end{equation}
\begin{equation}
\frac{\partial \vec{r}}{\partial t} + \vec{\omega} \times \vec{r} - (\frac{\partial \vec{v}}{\partial s} + \vec{\kappa} \times \vec{v})  =\vec{0} \label{position_continuity}
\end{equation}
\begin{equation}
\frac{\partial \vec{\kappa}}{\partial t} - (\frac{\partial \vec{\omega}}{\partial s} + \vec{\kappa} \times \vec{\omega})  =\vec{0}  \label{orient_continuity}
\end{equation}

Here $s$ is the cross-section location along the rod, $t$ is time, $m(s)$ is the mass of the rod per unit length and tensor $\underline{\mathbf{I_m}}(s)$ is the moment of inertia per unit length in the body-fixed frame of reference. Variation of vector $\vec{r}(s, t)$ encodes shear and extension of the rod. In this paper, it is assumed constant to ensure in-extensibility and un-shearability. The vectors $\vec{F}$ and $\vec{Q}$ are the external distributed force and moment, respectively. They include the distributed follower force as well as interactions of the rod with the environment such as fluid drag. Note that the spatial and temporal derivatives in equations (\ref{linear_momentum}) - (\ref{orient_continuity}) are relative to the body-fixed frame, which obviates the need of transforming body-fixed follower forces and drag to inertial frame.

The unknown variables that we need to solve for are: the vector $\vec{\kappa}(s,t)$ that captures two-axes bending and torsion, the vectors $\vec{v}(s,t)$ and $\vec{\omega}(s,t)$ that represent the translational and the angular velocities of each cross-section, respectively, and the vector $\vec{f}(s,t)$ that represent internal shear force and tension. The internal moment vector $\vec{q}(s,t)$ in the angular momentum equation (\ref{angular_momentum}) is related to $\vec{\kappa}(s,t)$ by the linear constitutive law

\begin{equation}
\vec{q}(s,t) = \underline{\mathbf{B}} \vec{\kappa},  \label{const}
\end{equation}
where the tensor $\underline{\mathbf{B}}(s)$ represents the bending and torsional stiffness of the rod. In the body-fixed frame that coincides with principal torsion-flexure axes, the stiffness tensor $\underline{\mathbf{B}}$ is expressed as
\begin{eqnarray}
   [\underline{\mathbf{B}}]=
  \left[ {\begin{array}{ccc}
    EI_1 & 0 & 0 \\
    0 & EI_2 & 0 \\
    0 & 0 & GI_3\\
  \end{array} } \right],
  \label{stiffness}
\end{eqnarray}
where $E$ is the Young's modulus, $G$ is the shear modulus, and $I_1$, $I_2$, and $I_3$ 
are the second moments of cross-section area about the principal torsion-flexure axes.

%The distributed follower forces and moments in this model are captured by $\vec{F}$ and $\vec{Q}$. In the scenario of fixed-fixed or pinned-fixed rod, we consider the effect of distributed follower forces, $F$ in tangential direction, see Figures 2 and 4 in the Results Section.% (along $\hat{a}_3(s, t)$). Here forward for simplicity of notation, we refer to this tangential follower force density by scalar $F$. 

%The differential equations of equilibrium and compatibility (\ref{linear_momentum}) - (\ref{orient_continuity}) have to be solved together with a constitutive law relating the deformations to the restoring forces.  The constitutive law, for an isotropic and linearly elastic rod takes the form of an algebraic constraint, $ \vec{q}(s,t) = \underline{\mathbf{B}}(s) \vec{\kappa}(s,t)$ where matrix $\underline{\mathbf{B}}$ encodes the bending and torsional stiffness moduli of the rod.  

The \textit{Generalized-$\alpha$} method is adopted to compute
the numerical solution of this system, subjected to necessary
and sufficient initial and boundary conditions. A detailed description of 
this numerical scheme applied to the rod formulation is given in \cite{sfb}. We have validated this scheme by comparing our results with the known results of Beck's column \cite{sfb2}. 

%We explore how the frequency, $\omega(F, L_{\mathrm{ee}}/L)$, as well as critical load to initiate the oscillations, $F_{cr}$ depend on the pre-stress. A cylindrical rod with slenderness ratio of 800 is simulated.

To model fluid dissipation, we have used either Stokes -like linear drag [S] or the quadratic Morrison drag [M] in our simulations. These drags are given by equations (\ref{eq:dragS}) and (\ref{eq:dragM}) below, respectively \cite{sachin}:
\begin{equation}
\vec{F}_{\textrm{S}}=-\frac{1}{2}\rho_{\textrm{f}} d \Big( C_n \vec{t}\times(\vec{v}\times \vec{t}) + \pi C_t(\vec{v}\cdot \vec{t})\:\vec{t} \Big)
\label{eq:dragS}
\end{equation}
\begin{equation}
\vec{F}_{\textrm{M}}=-\frac{1}{2}\rho_{\textrm{f}} d \Big( C_n|\vec{v}\times \vec{t}|\vec{t}\times(\vec{v}\times \vec{t}) + \pi C_t(\vec{v}\cdot \vec{t})|\vec{v}\cdot \vec{t}|\:\vec{t} \Big)
\label{eq:dragM}
\end{equation}
Here, $\rho_{\textrm{f}}$ is the fluid density, $d$ is diameter of the rod, ${\bf t}$ is the unit tangent vector along the rod's centerline and $C_n$ and $C_t$ are drag coefficients in the normal and tangential directions, respectively. In this paper we primarily focus on results obtained for quadratic drag. However, we also comment on results obtained using the linear drag.

\begin{table*}[htp]
\begin{center}
{\small 
   \begin{tabular}{ | l | l | l | l |}
    \hline
    {\bf Quantity} & {\bf Variable} & {\bf Value} & {\bf Units} \\ \hline\hline
    Diameter & $d$ & $ 0.0096 $ & m \\ \hline
    Length & $L$ & 8 & m \\ \hline
    Mass per unit length& $m$ & 0.2019 & kg/m \\ \hline
    Young's modulus & $E$ & 68.95 & GPa \\ \hline
    Shear modulus & $G$ & 27.58 & GPa \\ \hline
    Second moment of area & $I_1=I_2=I$ & 4.24 $\times$10$^{-10}$ & m$^4$ \\ \hline
    Polar moment of area & $I_3$ & 8.48 $\times$10$^{-10}$ & m$^4$ \\ \hline
    Normal drag coefficient & $C_n$ & 0.1 & s/m [S], none [M] \\ \hline
    Tangential drag coefficient & $C_t$ & 0.01 & s/m [S], none [M] \\ \hline
    Surrounding fluid density & $\rho_{\textrm{f}} $ & 1000 & kg/m$^3$  \\ \hline
    \end{tabular}
    \caption{Numerical values for the geometric and elastic properties of the rod and drag coefficients used in the simulations. {The ratio $C_n/\pi C_t = 3.18$ is comparable to the value 2 for the limit of purely viscous (Stokesian) drag ratio for a slender rod using resistivity theory} \cite{chelakkot}.}
    \label{tab1}}
\end{center}
\end{table*}

\section{Results}
\label{sec:3}

In this section, we present and compare the simulation results for the post-buckling analysis of pre-stressed rods with fixed-fixed (FF) and pinned-fixed (PF) boundary conditions. The aim is to quantify the effect of pre-stress on the stability margin for both of these boundary conditions. We also compare the results with the cantilever (fixed-free) scenario to shed more light on how pre-stress can be used to manipulate the onset of oscillations (i.e., the critical point).

\begin{figure}
\begin{center}
  \includegraphics[width=0.66\columnwidth]{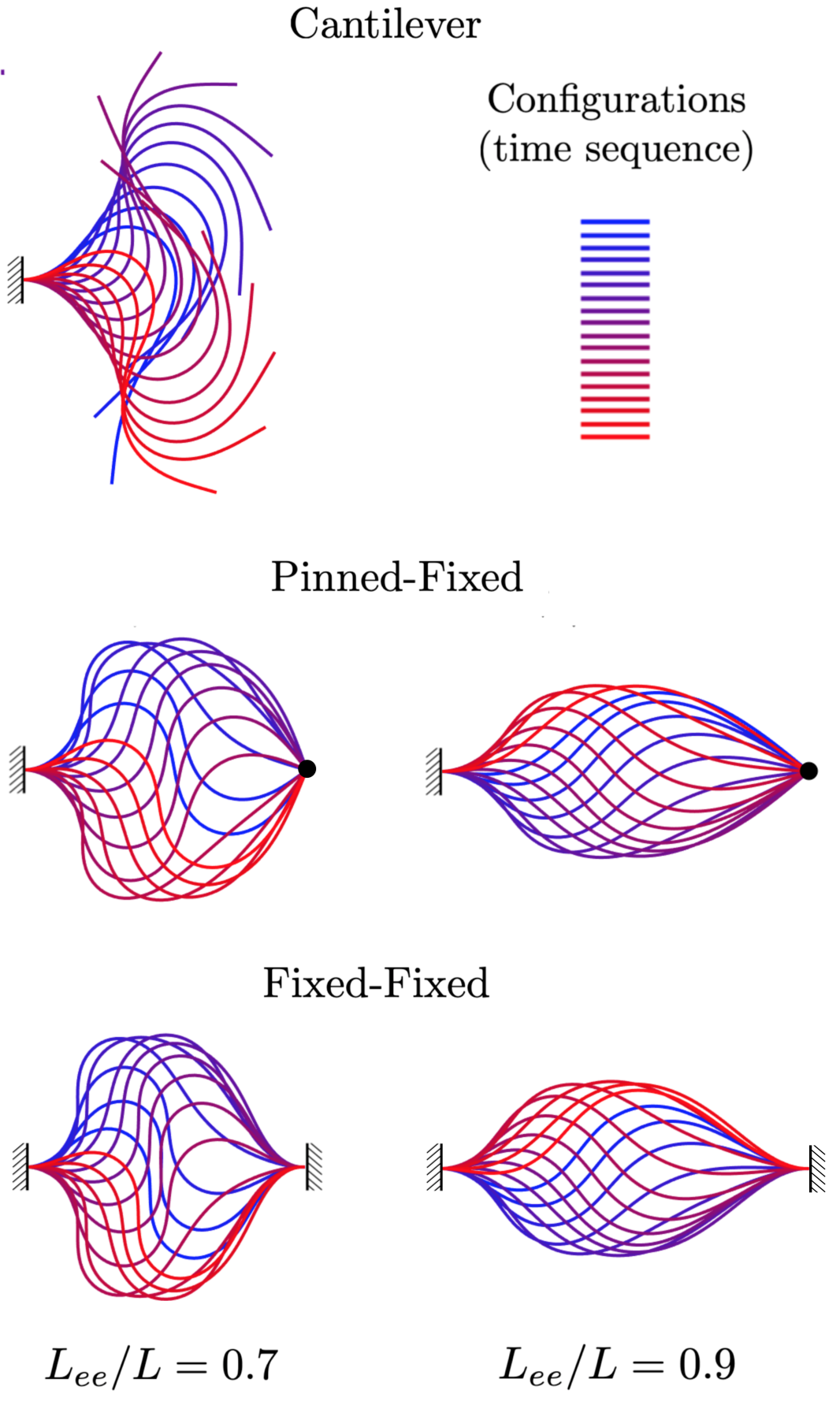}
\caption{Configurations of the oscillating rods are depicted during one time-period when $F=20$ N/m, and the drag
force is quadratic [M] in the rod velocity.}
\label{fig:3a}      
\end{center}
\end{figure}

In all the simulations an initially straight cylindrical rod is used with the properties given in Table 1. The pre-stress is generated by moving one end of the rod relative to and towards the other as shown in Figure 1. The pre-stress values are determined and controlled by the end-to-end distance, $L_{ee}$. Then we apply uniformly distributed follower load, $F$ to the pre-stressed rod along the tangential direction of the rod's centerline. As the follower force exceeds a critical value $F_{cr}$, the buckled equilibrium is destabilized and flapping oscillations emerge. The simulation snapshots in Figure 2 show some examples of how the shape of the rod evolves during one complete oscillation for all three boundary conditions (FF, PF and Cantilever) and $F=20$ N/m $>F_{cr}$. For the pinned-fixed scenario, flapping oscillations emerge only when the follower force points from the pinned end towards the fixed end. Upon reversing the direction of the follower force, flapping oscillations disappear and stable equilibria evolve. The stable equilibrium shapes that evolve in this scenario with increasing follower force are shown in Figure 3. For the cantilever, the follower force causes flapping oscillations when it points from free end towards fixed end, not otherwise. This is expected as instabilities are due to compressive stresses. Tensile stresses do not lead to instabilities.

%In both fixed-fixed and pinned-fixed scenarios starting with a straight rod we move one end of the rod towards the other. This process generates pre-stress in the rod as shown in Figure 1. Hence, pre-stress values are determined and controlled by the end-to-end distance, $L_{ee}$ as shown in Figure 1. 

\begin{figure}[t]
\begin{center}
  \includegraphics[width=0.92\columnwidth]{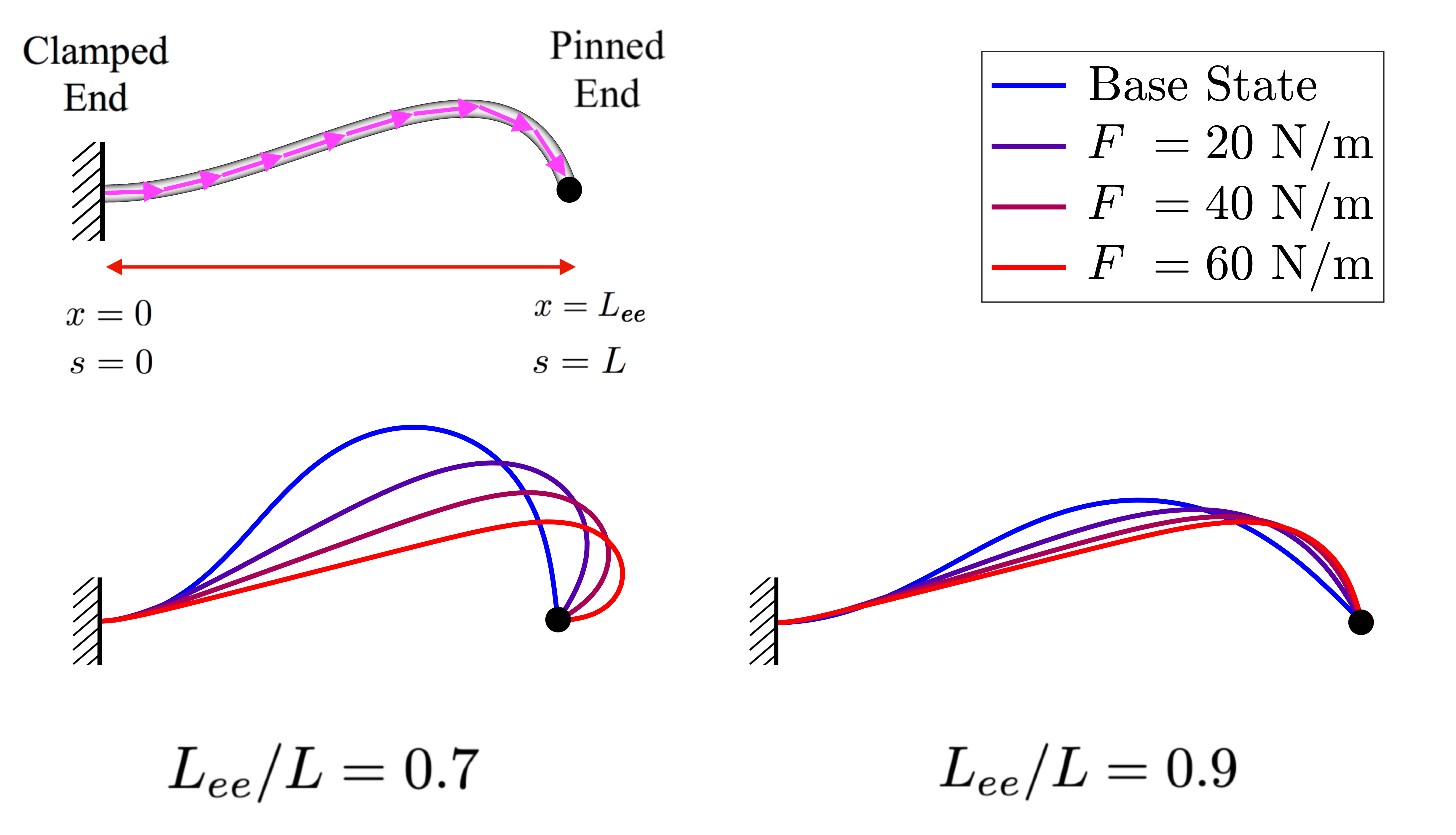}
\caption{In pinned-fixed scenario if the direction of the follower force is from the fixed end towards the pinned end, flapping oscillations do not occur. Instead the stable equilibrium shape keeps evolving with increasing follower force. }
\label{fig:3b}   
\end{center}    
\end{figure}

In the next sub-section, we analyze the flapping oscillations by looking at the transfer of energy to and from the rod, and the variation of total energy stored in the form of strain and kinetic energies. This allows us to rationalize how the instability-driven flapping oscillations are sustained in steady state. %We also present the condition under which no flapping oscillation is possible in pinned-fixed scenarios. 
Then, we present results for the critical value of the follower force $F_{\mathrm{cr}}$ versus pre-stress measured by end-to-end distance $L_{\mathrm{ee}}/L$ for both fixed-fixed and pinned-fixed scenarios. Next, we examine how the planar beating frequency, $\omega(F, L_{\mathrm{ee}}/L)$ both at the critical point and for values of the follower force $F > F_{\mathrm{cr}}$ depends on the pre-stress. To highlight the effect of pre-stress we also report $F_{\mathrm{cr}}$ of a cantilever (fixed-free) rod, which has no pre-stress, and show its frequency response as well. Finally, we discuss design implications of some of the observations.%In all cases, a cylindrical rod with slenderness ratio of 800 is simulated with properties given in Table \ref{tab1}. 

\subsection{Energy Exchange During Flapping}
\label{sec:3_0}

Figure 4 shows how strain energy, kinetic energy, work done by follower force, and energy dissipated by fluid drag evolve as flapping oscillations emerge and reach a steady state in a pinned-fixed scenario for $F=F_{cr}$, and slack (related to end-to-end distance) $1 - L_{ee}/L$ $= 0.3$. Snapshots of the shape show the rod flapping from the base state (A) to (C) via strain energy maxima (B) and symmetrically flapping back to configuration (E) via (D). In each cycle, the follower force does work to (i) increase the elastic energy stored in the rod via the strain field, (ii) increase the kinetic energy of the rod, and (iii) overcome the fluid dissipation (from A to B or C to D). Specially, we note the steep ramp-up in these intervals corresponding to an increase in strain energy. In contrast, between (B) and (C), or (D) and (E), the total mechanical energy stored in the system (strain energy and kinetic energy) continues to drive the oscillations overcoming the negative work done by follower force and again the fluid dissipation. Thus elastic and kinetic energies of the deforming rod serve to mediate the transfer of energy eventually from the active forces to the ambient fluid in each cycle. 

%We also observe a concurrent increase in net energy dissipation and kinetic energy that is due to the quadratic dependence of drag force on velocity. Moreover, in
When the rod shapes from (A) through (E)--the complete cycle--are superimposed (left top graphic in Figure 4), it is obvious that while flapping back the rod doesn't retrace its configuration. The graphic also shows trajectories of three points at $s=0.25L$, $s=0.50L$, and $s=0.75L$ on the rod. The points don't retrace their paths, but instead follow a figure 8-like loop. If the entire rod were to retrace back its configuration while flapping back, i.e., if all points were to retrace their paths, the follower force could not have done any net work.

\begin{figure*}[t]
\begin{center}
  \includegraphics[width=1.88\columnwidth]{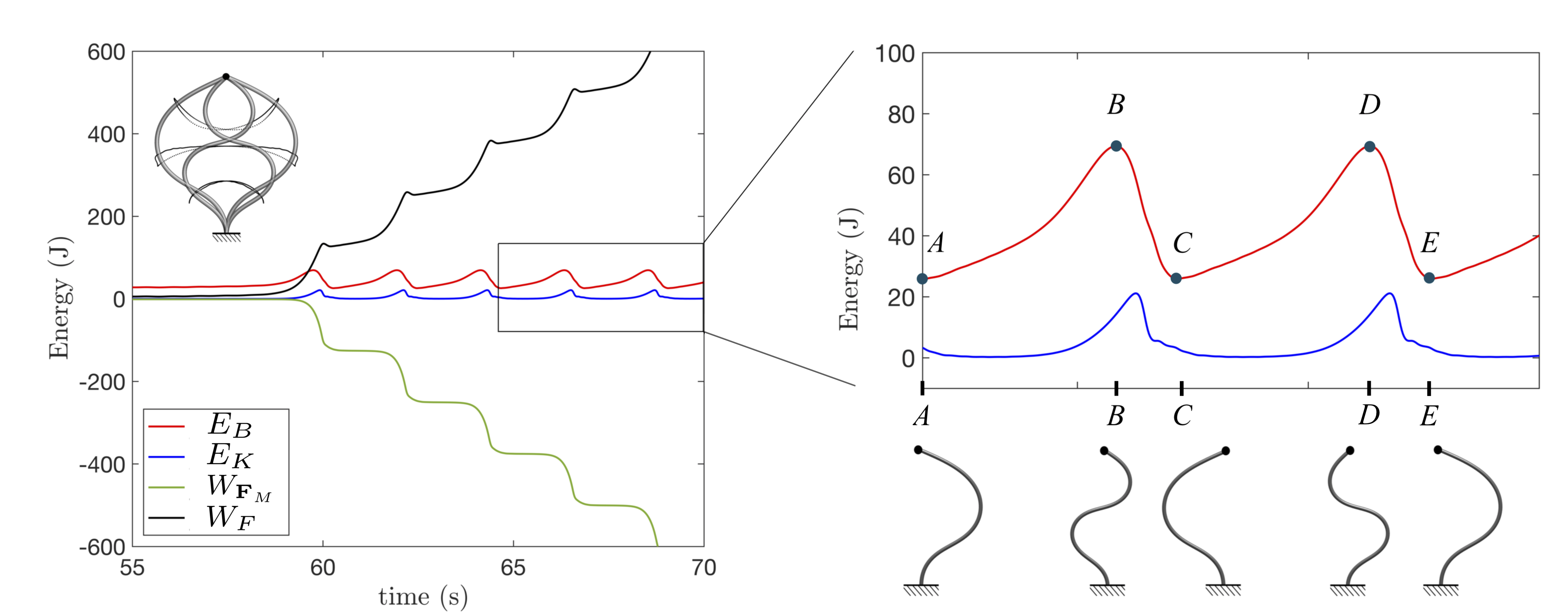}
\caption{Tracking strain energy, $E_B(t)= \int^L_0 \frac{1}{2}E\underline{\mathbf{I_m}}\vec \kappa \cdot \vec \kappa ds$, kinetic energy, $E_K(t)= \int^L_0 \frac{1}{2}m\vec v \cdot \vec v ds$, for $F=F_{cr}$ and slack $1 - L_{ee}/L=0.3$, dissipation energy, $W_{{\bf F}_M}(t)= \int^t_0 \int^L_0 {\bf F}_M \cdot \vec v ds d\tau $, and work done by follower force, $W_{F}(t)= \int^t_0 \int^L_0 F\vec t \cdot \vec v ds d\tau $ illustrates that intervals of positive and negative work done by follower force correspond to increase and decrease in strain energy, respectively, and that intervals of peak kinetic energy correspond to jumps in energy dissipation. Points on the rod are found to trace an 8-like shaped loop. On the right, strain energy, kinetic energy, as well as the snapshots of the rod shapes at the base state energy level (A, C, and E) and at the maximum of strain energy (B and D) show the exchange of energy during oscillations. }
\label{fig:4}   
\end{center}    
\end{figure*}

\subsection{Critical Force for Onset of Flapping}
\label{sec:3_1}

\begin{figure*} [t] 
\begin{center}
  \includegraphics[width=1.72\columnwidth]{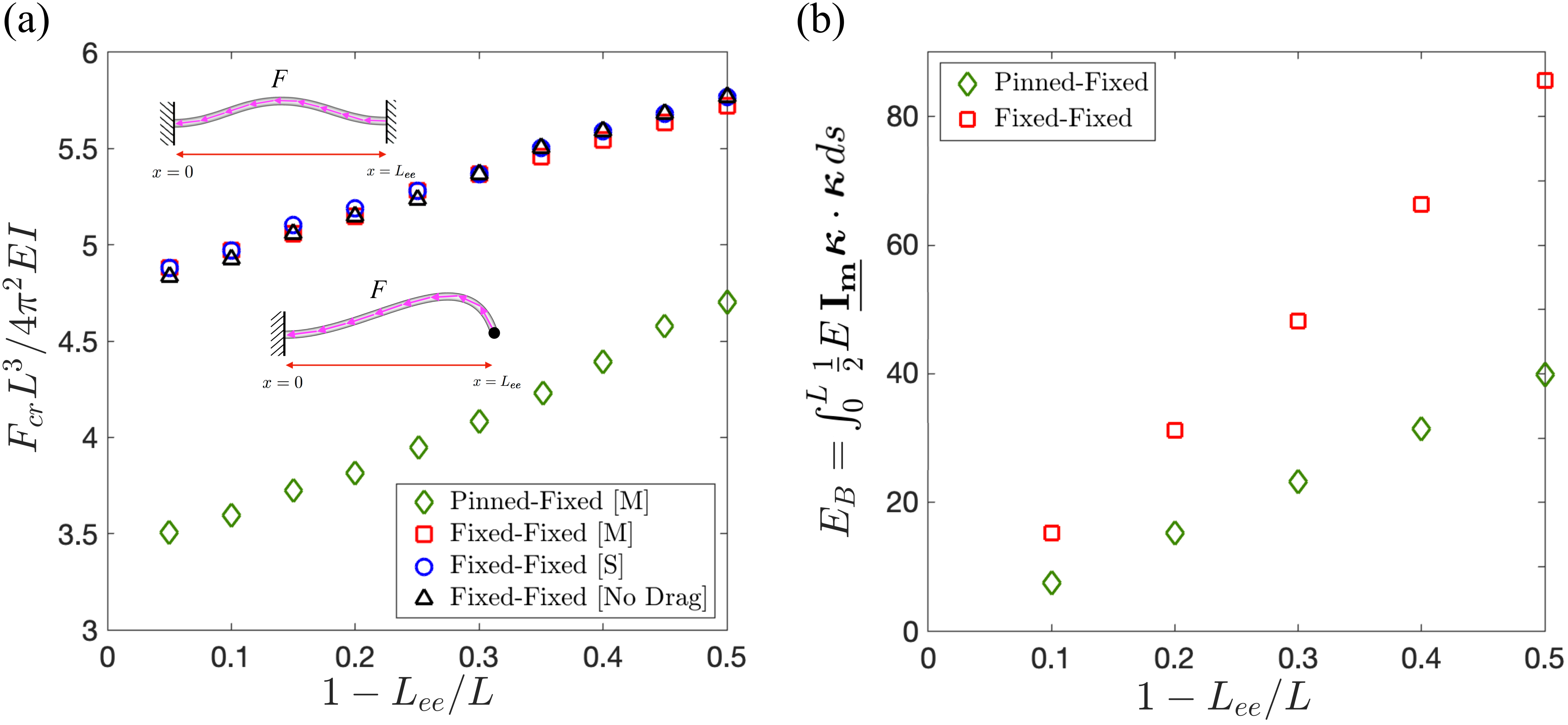}
\caption{(a) Critical load for onset of oscillations, $F_{\mathrm{cr}}$ versus normalized decrease in end-to-end distance, $1- L_{\mathrm{ee}}/L$ for both pinned-fixed and fixed-fixed scenarios.
For $0.05 <1- L_{\mathrm{ee}}/L < 0.5$, the critical force $F_{\mathrm{cr}}$ increases as $1- L_{\mathrm{ee}}/L$, or pre-stress, increases. Normalized critical load for stress-free cantilever is $F_{\mathrm{cr}}L^3/(4\pi^2 EI)=0.0916$  (b) Strain energy of the base state, $E_B$ increases with pre-stress. It is also higher for fixed-fixed boundary condition than for the pinned-fixed condition.}
\label{fig:2}       
\end{center}
\end{figure*}

The critical values of the distributed follower force (or critical follower force densities) are computed by numerically integrating the time dependent equations (1)-(7) and seeking the point at which stable oscillations emerge as the follower load increases. Note that this is done using time-integration and not via continuation methods. This procedure is repeated for several values of pre-stress, $1-L_{\mathrm{ee}}/L$. Since our aim is the identification of the parameter range that can be explored experimentally, we focus only on the pre-stress values satisfying $0.95 < 1- L_{\mathrm{ee}}/L $. 

As soon as the magnitude of the follower load is above the critical value, $F>F_{\mathrm{cr}}$, base states become unstable and oscillations emerge. Figure 5(a) shows the magnitudes of the critical follower load $F_{\mathrm{cr}}$ against the slack, $1-L_{\mathrm{ee}}/L$ for both fixed-fixed (FF) and pinned-fixed (PF). We find that (i) in both cases the critical follower force density increases in magnitude as the pre-stress in the rod increases (ii) for the same end-to-end distance, other things being equal, FF boundary condition has a larger critical point in comparison to PF, and (iii) the magnitude of critical follower load is nearly the same for linear drag, quadratic drag, or no fluid drag (discrepancies being $<2$\%). We can explain the first finding by looking at the strain energies of the base states shown in Figure 5(b). Larger slack corresponds to larger pre-stress which in turn implies that base state has a larger strain energy. Hence, as slack increases a larger follower force is required to overcome the larger barrier of elastic energy in order to initiate the flapping oscillations. Similarly, the second finding can also be explained by the fact that for a given pre-stress value, FF base states possess higher strain energy than do the PF base states. And finally, the third finding can be explained by the fact that critical point is governed by the linear stability of the system while undergoing small perturbations, therefore nonlinear [M] and linear [S] drags yield the same critical value. In addition, we surmise that the onset of oscillations and the onset of flutter are very close to one another for the parameter range investigated here since values of critical load in absence of drag are close to the values found with fluid dissipation.

Finally, % in Figure 5(a), we observe that for a given end-to-end distance, FF boundary condition results in a larger critical load than the PF condition. Nonetheless,
by examining the slopes in Figure 5(a), we find that the rate at which the critical values increase with pre-stress is larger for PF than for FF, hence PF stability region is more sensitive to pre-stress than FF stability region. 

\subsection{Frequency of Flapping}
\label{sec:3_2}

We next examine the frequency of oscillations, i.e., the wave speed associated with the propagation of curvature along the arc-length as the rods cycle in configuration -time space. In all simulations we observe that for $F \geq F_{\mathrm{cr}}$ oscillations eventually reach a stable state, implying the rate of energy input into the system due to the action of the nonconservative follower forces balances the rate of energy dissipated by the fluid drag. We track the oscillations for 40 seconds for each $F$ value corresponding to a minimum of 8 full oscillations up to a maximum of 70 full oscillations (once stable state is attained).

The computational model described in section 2 is used here to systematically investigate the effect of pre-stress and the follower force on the frequency of oscillations, $\omega(F, L_{\mathrm{ee}}/L)$ near the critical point as well as far from the critical point where $|F - F_{\mathrm{cr}}|/F_{\mathrm{cr}} > 1$. To better understand the results, we juxtapose the cantilever (stress-free) force-frequency curve with those of fixed-fixed (FF), and pinned-fixed (PF) loading scenarios. Figure 6 illustrates the frequency of flapping oscillations for rods under various end-to-end distances and subjected to Morrison drag. The frequency values are cast on a log-log scale; shown alongside is the power law relating the frequency to the follower force, $\omega \sim F^{5 \over 6}$ found theoretically using scaling arguments based on power and dissipation rates \cite{sfb2}. When oscillations reach steady state the rate at which energy enters the system due to the work done by nonconservative follower force balances the rate at which energy dissipates due to the fluid drag. 

Figure 6 illustrates that frequencies in both PF and FF cases converge to that of the cantilever as the pre-stress vanishes. Moreover, far from the critical point, it can be observed that force-frequency curves for all three loading scenarios and all pre-stress values collapse into one. 

\subsection{Implications for Design}

In this section we discuss some of the implications of this study for design of synthetic active filaments in applications such as biomimetic soft robots or micro fluidic devices. First, we comment on how pre-stress can be potentially used to control the oscillations, second we discuss the role boundary conditions can play in regulating the dynamics of oscillations. All of these implications are grounded on, and hence are relevant to the range of parameters explored in this paper.%, however, they need not be necessarily invalid for situations unexamined here.

\subsubsection{Regulation by pre-stress} 

Based on the results presented in Figure 5, we conclude that pre-stress can be effectively used to regulate the onset of oscillations and to control the stability margin in both FF and PF cases. In the region near the critical point, pre-stress can be also an effective parameter to control the frequency of the oscillations in both FF and PF scenarios, however far form the critical point the frequency of oscillations become independent of pre-stress and independent of the boundary conditions.

\subsubsection{Regulation from boundary constraints}

Another important observation we have made is that in PF scenario if the direction of the follower force is from the fixed end towards the pinned end (F to P), there would be no emergent oscillations possible. Thus, all the PF results presented here can be produced only when direction of the follower force is from the pinned end towards the fixed end (P to F). This feature can be used to manipulate the oscillations. Altering boundary conditions independently from the mechanism that generate follower force can be used to start or stop the flapping oscillations. For example, in FF scenarios by converting one clamped end into a pin joint this feature can be used to suppress oscillations. %For example, in FF scenarios by switching one of the clamped end--the one towards which the follower force is directed--into a pin joint the oscillations can be suppressed. %Figure 2 illustrates the rod shapes in PF scenario subjected to a follower force directed from the fixed end towards the pinned end and after the structure reaches static equilibrium (40 seconds per each loading step).

\begin{figure}
  \includegraphics[width=8cm]{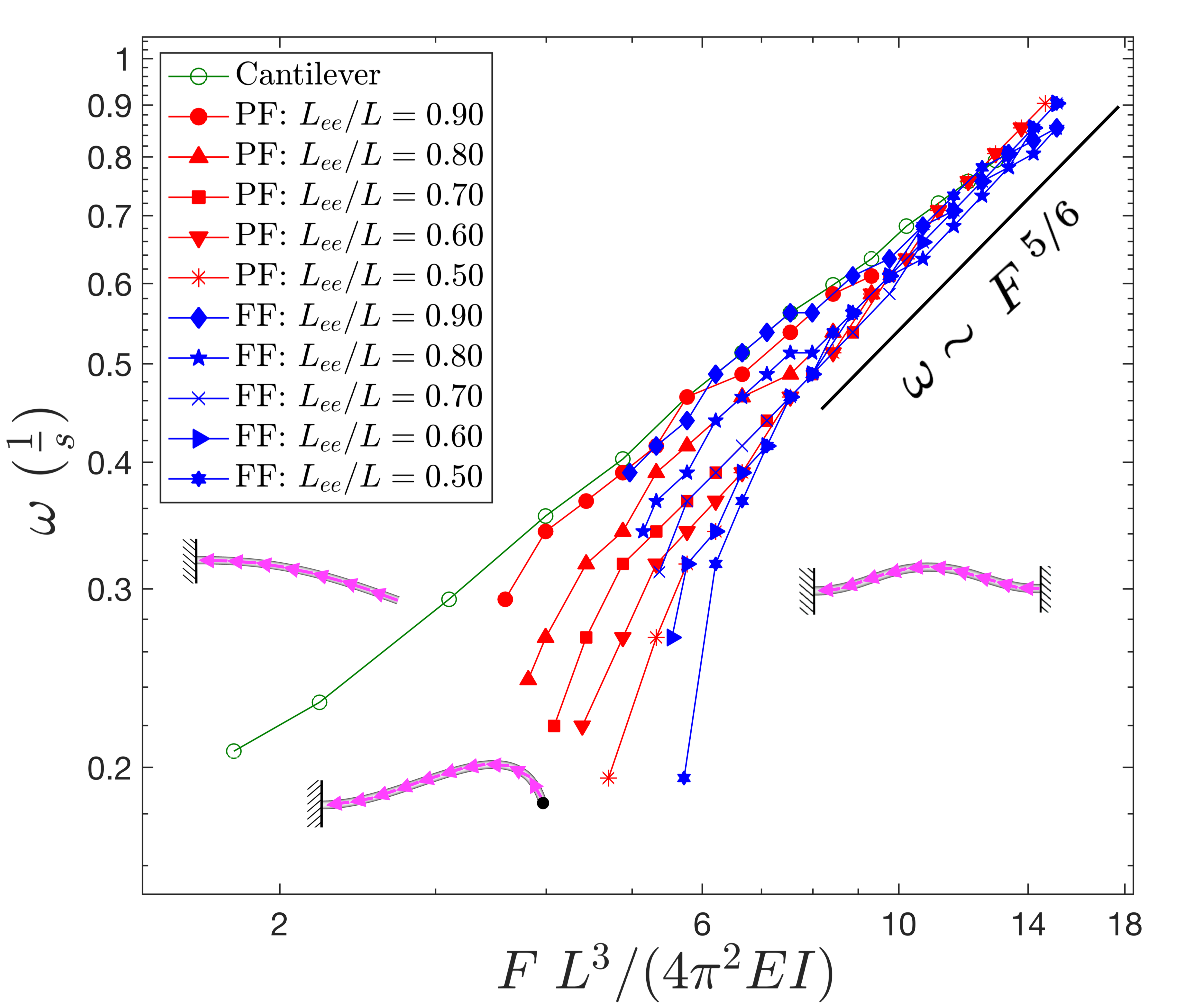}
\caption{Frequency for the Morrison [M] drag plotted as a function of the force density $F$ plotted in logarithmic scales to illustrate two salient features; (i) as the follower force increases to values much larger than the critical values, the effect of the pre-stress diminishes--similar frequencies are observed for fixed-fixed (FF) and pinned-fixed (PF) scenarios far from criticality--and (ii) the frequencies in the limit $F \gg F_{\mathrm{cr}}$ scale roughly as $\omega \sim F^{5 / 6}$ consistent with our theoretical prediction.}
\label{fig:5}      
\end{figure}

\section{Conclusions}
\label{sec:4}

In this paper we analyzed the post-buckling flapping oscillations of constrained slender structures subjected to tangential follower loads using a computational rod model. This scheme was benchmarked by comparing with established results of critical buckling force for Beck's column \cite{sfb2}. We focused on slender rods maintaining a straight shape in stress-free state (i.e., having neither intrinsic curvature and twist nor axial tension) with boundary conditions being either both ends clamped, or one end clamped and the other pinned. By moving the two ends of the rod towards one another, the structure is forced to buckle. Thus, effects of pre-stress and boundary conditions are systematically tested both on the emergence of buckling instabilities, as well as on the post-buckling oscillations induced by follower force. In these computations the inertia of the rod, geometry and the fluid drag coefficients are held fixed. We found that beyond a critical value of distributed and compressive follower loads the buckled shapes become unstable and oscillatory beating emerges. This critical value is found to be larger for rods with fixed-fixed boundary condition in comparison to the rods with pinned-free constraints. Nonetheless, the magnitude of the critical follower load increases as the magnitude of the pre-stress in the structure increases. 
Far from criticality, i.e. for $F$ much greater than the critical value needed to initiate the oscillations, the response frequencies exhibit a power law dependence on $F$ with an exponent $5\over 6$. This exponent is explained by consideration of a power balance between the active energy pumped into the system by the nonconservative follower forces and energy dissipated due to fluid drag.

As mentioned earlier, we have found critical forces for pinned-fixed condition to be smaller that the critical force for fixed-fixed boundary condition for the same value of the slack. This is consistent with previous work on animated filaments without pre-stress \cite{chelakkot} where it is found that the critical load for pinned-free scenario is smaller than that of fixed-free (cantilever). A linear stability analysis for these two cases indicates that non-trivial solutions emerge from the trivial state via a Hopf-Poincare bifurcation with flapping (complex conjugate eigenvalues crossing the real axis) for fixed-free loading condition. For the pinned-free scenario, since there is a rotational degree of freedom at the pin and no energy penalty for free-rotations about this point, the linear stability suggests a simple global bifurcation (single eigenvalue crossing zero) and the nonlinear stable state is a rotating coil. In our study, when the follower force is directed towards the pinned end while the other end is clamped, the strained rod cannot rotate about the pin, instead it deforms and reaches a state of static equilibrium; in the vicinity of the pivot the rod is highly curved. 
 
Our approach provides a platform to investigate the interplay between geometry, elasticity, dissipation, and activity and overall to contribute towards designing bio-inspired multi-functional, and synthetic structures to manipulate and control fluid at various
length scales or generate propulsion in soft robotics. Further extensions and developments of this study need to examine the stability margin and dynamics of emergent oscillations subjected to three-dimensional perturbations. Moreover, the fluid-structure interaction model can be improved to incorporate two-way coupling and hence analyze, inter alia, an ensemble of filaments and their interaction. Finally, continuation and homotopy methods using Newton-GMRES  \cite{Anwar} or variants that are adapted to use time-steppers to trace both unstable and stable solutions branches will complement the analysis presented here.

%\begin{table}
%\begin{tabular}{lll}
%\hline\noalign{\smallskip}
%\end{tabular}
%\end{table}
\vspace{0.5cm}

\textbf{Conflict of Interest}: The authors declare that they have no conflict of interest.

%\begin{acknowledgements}
%\end{acknowledgements}



\end{document}